\documentstyle[12pt]{article}
\def\be{\begin{equation}}
\def\ee{\end{equation}}
\def\ba{\begin{eqnarray}}
\def\ea{\end{eqnarray}}
\def\ga{\mathrel{\raise.3ex\hbox{$>$\kern-.75em\lower1ex\hbox{$\sim$}}}}
\def\la{\mathrel{\raise.3ex\hbox{$<$\kern-.75em\lower1ex\hbox{$\sim$}}}}
\newcommand{\sect}[1]{\section{#1}\setcounter{equation}{0}}

\def\theequation{\thesection.\arabic{equation}}

\newcommand{\bi}[1]{\bibitem{#1}}

\begin{document}
\baselineskip=16pt
\begin{titlepage} 
\rightline{UMN--TH--1829/99}
\rightline{TPI--MINN--99/54}
\rightline{OUTP-99-62P} 
\rightline{hep-ph/9912266}
\rightline{December 1999}  
\begin{center}

\vspace{0.5cm}

\large {\bf Single-Brane Cosmological Solutions\\[2mm]
with a Stable Compact Extra Dimension}
\vspace*{5mm}
\normalsize

{\bf Panagiota Kanti$^1$, Ian I. Kogan$^2$, Keith A. Olive$^1$} and {\bf
Maxim Pospelov$^1$}

\smallskip 
\medskip 
 
$^1${\it Theoretical Physics Institute, School of Physics and
Astronomy,\\  University of Minnesota, Minneapolis, MN 55455, USA} 

$^2${\it Theoretical Physics, Department of Physics, Oxford University}\\
{\it 1 Keble Road, Oxford, OX1 3NP,  UK}
\smallskip 
\end{center} 
\vskip0.6in 
 
\centerline{\large\bf Abstract}

We consider 5-dimensional cosmological solutions
of a single brane. The correct cosmology on the brane, i.e., 
governed by the standard 4-dimensional Friedmann equation, and stable
compactification of the extra dimension is
guaranteed by the existence of a non-vanishing $\hat{T}^5_5$ which is
proportional to the 4-dimensional trace of the energy-momentum tensor.
We show that this component of the energy-momentum tensor arises from the
backreaction of the dilaton coupling to the brane.
The same positive features are
exhibited in solutions found in the presence of non-vanishing
cosmological constants both on the brane ($\Lambda_{br}$) and in the bulk
($\Lambda_B$). Moreover, the restoration of the Friedmann equation, with
the correct sign, takes place for both signs of $\Lambda_B$ so long as the
sign of $\Lambda_{br}$ is opposite $\Lambda_B$ in order to cancel
the energy densities of the two cosmological constants.
We further extend our single-brane thin-wall solution to allow a brane
with finite thickness.

\vspace*{2mm} 

\end{titlepage} 

\sect{Introduction}

It goes without saying that there has been an exceptional amount of
interest in theories with extra dimensions, and in particular the
resulting cosmologies that arise in those theories.  In the past,
in theories with extra dimensions, one often simply assumed a small
compact extra dimension, of order the Planck size. In this case, it is
usually sufficient to average over the extra dimension(s) and
once fixed, cosmology on the large 4D space-time is conventional, other
than the appearance of additional massive degrees of freedom. 
In 10D string theory, while the structure of the compact 6D space has a
profound effect on the resulting interactions of matter in the effective
low-energy theory, cosmology is only affected by terms suppressed by
inverse powers of the Planck scale. Below the Planck scale, these are
generally negligible. When string theory is promoted to M-theory, the
size of the extra 11th dimension, is generally assumed to be somewhat
larger than $M_P^{-1}$, and thus allows for a separation of the string
(unification) scale and the Planck scale \cite{w}.

By allowing the extra dimension to be significantly larger than the
Planck scale, one can try to relate the electroweak scale to the
fundamental higher dimensional Planck scale \cite{large,D1,D2}.
Alternatively, it has been suggested that an exponential scaling of the 
metric along the extra dimension (or ``warp'' factor) can also lead to a
resolution of the hierarchy problem
\cite{RS}.  In both of these approaches the extra dimension may be as
large as a mm scale (in the former) and even infinite (in the latter,
see also \cite{RS2}).   In these theories, it is necessary to assume that
standard model particles are confined to a 3+1-dimensional brane of
an $n$+3+1-dimensional space-time. On the other hand, gravity
propagates in the bulk. 

Many theories have been constructed using the notion that
there are in fact two branes which represent the boundaries of a higher
dimensional space-time. Such was the motivation from the Horava-Witten
model in M-theory \cite{w}. Indeed, there has been considerable effort
to produce cosmological models on such boundaries \cite{dhlmp,LOW,LK}.  
In a different set of solutions \cite{RS}, the brane is a slice of AdS
and the negative cosmological constant of the bulk is used to cancel the
cosmological constant or tension on the brane. In the single brane
scenario \cite{RS2}, gravity is effectively confined to the brane by the
steep warp factor generated by the tension dominating the brane. 
In most other constructions, two branes are necessary to achieve
compactification. Of course in realistic cosmologies, the energy density
of the Universe must be dominated by matter. 

An important observation about the cosmology of  the 
``brane world'' was made in Refs. 
\cite{LOW,LK,BDL}. These papers showed that the Hubble
parameter $H$ governing the expansion of the scale factor on the brane
has a different behavior than derived from the usual 4-d Friedmann
equations. In particular, the Hubble parameter is
proportional to  the energy density on the brane instead of the familiar
dependence $H\sim 
\sqrt \rho$. While mathematically correct as a solution to Einstein's
equations, this behavior, if nothing else, indicates that some key
ingredient is missing if this theory is to be capable of describing our
Universe. It is worth noting  that this unusual behavior of
$H$ does  not depend on the size of the extra dimension, but holds even for
$r\sim  M_*^{-1}\sim M_{\rm Pl}^{-1}$. 
While this abnormal behavior of $H$ may be possible at very 
early cosmological epochs which cannot be immediately confronted 
with observational data, the $H\sim\rho$ -type of behavior is 
totally unacceptable at later times, 
particularly at big bang nucleosynthesis. 

The origin of the aberrant expansion law can be traced to a condition,
imposed on the solution of the 5D Einstein's equations, which relates the
energy density on the two branes. Roughly speaking, the presence of
matter on the 3-brane, produces a $y$-dependence ($y$ is the spatial
coordinate of the extra dimension) of the 3-space scale factor $a$ such
that $a(y)$ decreases as one moves away from the brane. Indeed for an
infinitely thin brane, there is a discontinuity in the derivative of $a$
with respect to $y$. (This discontinuity is removed by considering branes
of finite thickness, but the qualitative behavior of the solutions are
unchanged as we will show below.) For a brane dominated by a positive
tension, $a(y)$ decreases exponentially away from the brane, and 
in effect compactification is unnecessary \cite{RS2}. If the brane is
not dominated by the tension, the decrease in $a(y)$ is softer and
compactification may be necessary. To compactify, a second brane must be
placed  at some fixed distance from the initial brane. The scale factor
will also experience a discontinuity (in $a^\prime(y)$) on the second
brane, but of opposite sign.  This distinction imposes very severe
constraints on the matter and tension of the second brane.  In fact, if
the energy density on the original brane is $\rho_1$, then on the second
brane it must be $\rho_2 = -\rho_1 + O(\rho^2)$. At cosmological distance
scales (much larger than the separation of the branes), the Hubble
parameter will be given by $H^2 M_P^2 = \rho_1 + \rho_2 \sim \rho^2$. 
This point was recently emphasized in \cite{shirom}.

A number of ``remedies'' to this problem have been suggested
\cite{NK,Berkl,Cline,Kor}. Most of these works exploit 
the Randall-Sundrum construction with 
two branes of opposite tensions $\pm \Lambda_{br}$
and obtain the correct cosmological Friedmann equation by
cancelling the leading $\Lambda_{br}^2$-proportional 
term  with a negative bulk 
cosmological constant $\Lambda_B$. The remaining cross-term leads to the 
normal expansion law and the 4D Plank scale is related to the tension  on
the brane. 
For an alternative solution to this problem see \cite{FTW}. 
Cosmology of the single-brane scenario of ref. \cite{RS2} has
been considered in \cite{Berkl,Ida}. For additional papers of
interest related to the Randall-Sundrum scenario see
\cite{moreref}.
  
In a different approach to the problem \cite{kkop1} we  derived 
sufficient conditions which ensure
a smooth transition to conventional cosmology and Newton's law on the
brane. We found that if  $T_{\mu\nu}^{bulk} \neq 0$, and in particular,
$T_{55} \ne 0$ on the brane and in the bulk, the transition to the
conventional cosmology can be obtained (for both thin and thick branes).
Furthermore, the same class of solutions \cite{kkop1}
allowed for the possibility of compactification with {\em single} brane,
and obviates the need for an exotic object such as a brane with negative
matter density.

The main result achieved in \cite{kkop1} 
is that the desired resulting expansion law ($H^2 \propto \rho$),
together with a fixed size ($L=const$) for the extra dimension can be
achieved when $T_{55} \ne 0$.
In fact, the specific value of $T^5_5$ must be chosen to be proportional
to the trace of the  4D energy-momentum tensor on the brane and
inversely proportional to the size of an extra dimension, $T^5_5 \propto
(-\rho+3p)/2L+ O(\rho^2)$.  It was hypothesized in  \cite{kkop1} that this
value is in fact the consequence of the physics responsible for the 
stabilization of the dilaton. 

In this paper we take this idea one step further and 
study the relation between $T^5_5$ and the dilaton fixing in more detail.
In the absence of the dilaton stabilization mechanism certain 
conditions should be imposed on the 5D energy-momentum tensor 
to preserve dilaton flatness and thus the stability of the radius of an
extra dimension. Working to linear order in energy density (assumed
small), we find a generic integral condition on the 
energy-momentum tensor of 5D matter.
When the explicit dilaton potential is introduced, we show that 
$T^5_5 \propto (-\rho+3p)/2L+ O(\rho^2)$ arises as a back-reaction of the
dilaton potential in the presence of the brane and, to this order, 
does not depend on the details of the dilaton fixing potential. That is,
the solution found in \cite{kkop1} is completely generic.

Another important goal that we pursue in this paper is the 
development of our initial solution \cite{kkop1}
in several new directions. In particular, we study whether the 
thin wall solution with a compact extra dimension allows 
for the generalization on the case of small but finite thickness. 
We also extend our solution to include cosmological 
constants on the brane and in the bulk, without restricting the sign of the 
bulk cosmological constant. In all of these cases we  recover normal
Friedmann-type solutions on the brane by requiring the stability of the 
transverse dimension. 

We organize this paper as follows. The main ingredients of the model,
Einstein equations specialized for particular
ansatzes for the metric and energy-momentum tensor, are introduced 
in the next section. Section 3 deals with
the interpretation of the (55)-component of the energy-momentum tensor in
the bulk. First, we consider the condition for dilaton flatness
in the presence of 5D matter and show the desirable form of 
$T_5^5$ which satisfies this condition. Then we turn to the most
important  case when the dilaton receives an explicit potential
and demonstrate how $T_5^5$ is generated as the 
response of this potential to the presence of the brane. 
In section 4, the thin wall 
solution is generalized to take into account cosmological 
constants on the brane and in the bulk. We address the cases of negative, 
positive and zero cosmological constant in the bulk. 
In section 5 the single brane solution with compactification is analyzed
when the brane is given a certain thickness $\Delta\ll L$.
Our conclusions are presented in section 6.

\sect{The Theoretical Framework}

As an example of a higher-dimensional theory describing the coupling of the
matter content of the universe with gravity, we consider the following
5-dimensional theory 
\be
S=\int d^5x \sqrt{-G^{(5)}}\,\Bigl\{\frac{M_5^3}{16\pi}\,\hat{R}
+ \hat{\cal L}_o \Bigr\}\,,
\label{action}
\ee
where $\hat{\cal L}_o$ represents all possible contributions to the action 
which
are not strictly gravi\-tational. In the above, $M_5$ is the fundamental
5-dimensional Planck mass, and the hat will denote 5-dimensional
quantities.  The line-element of the 5-dimensional manifold is given by
the following ansatz
\be
ds^2=-n^2(t,y) dt^2 + a^2(t,y) \delta_{ij} dx^i dx^j + b^2(t,y) dy^2\,,
\label{metric}
\ee
where $\{t,x^i\}$ and $y$ denote the usual, 4-dimensional spacetime 
and the extra dimension, respectively.

When the action functional (\ref{action}) is varied with respect to
the 5-dimensional metric tensor $G^{(5)}_{MN}$, Einstein's equations
are derived, which for the above spacetime background take the
form (see e.g. \cite{LOW,BDL})
\ba
\hat{G}_{00} &=& 3\Biggl\{\frac{\dot{a}}{a}\,\Biggl(\frac{\dot{a}}{a} +
\frac{\dot{b}}{b}\Biggr) -\frac{n^2}{b^2}\,\Biggl[\frac{a''}{a} +
\frac{a'}{a}\,\Biggl(\frac{a'}{a} - \frac{b'}{b}\Biggr)\Biggr]\Biggr\}
= \hat{\kappa}^2 \, \hat{T}_{00}\,,\label{00}\\[4mm] 
\hat{G}_{ii} &=& \frac{a^2}{b^2}\Biggl\{\frac{a'}{a}\,
\Biggl(\frac{a'}{a} + 2\frac{n'}{n}\Biggr) -\frac{b'}{b}\,\Biggl(\frac{n'}{n}
+2\frac{a'}{a}\Biggr) +2 \frac{a''}{a} +\frac{n''}{n}\Biggr\}\nonumber\\[4mm]
&+& \frac{a^2}{n^2}\Biggl\{\frac{\dot{a}}{a}\,\Biggl(-\frac{\dot{a}}{a} +
2\frac{\dot{n}}{n}\Biggr) -2\frac{\ddot{a}}{a}+ \frac{\dot{b}}{b}\,
\Biggl(-2\frac{\dot{a}}{a} + \frac{\dot{n}}{n}\Biggr) -
\frac{\ddot{b}}{b}\Biggr\}= \hat{\kappa}^2\,\hat{T}_{ii}\,,
\label{ii}\\[4mm]
\hat{G}_{05} &=& 3\Biggl(\frac{n'}{n} \frac{\dot{a}}{a}
+ \frac{a'}{a} \frac{\dot{b}}{b} -\frac{\dot{a}'}{a}\Biggr)=0\,,
\label{05}\\[4mm]
\hat{G}_{55} &=& 3\Biggl\{\frac{a'}{a}\,\Biggl(\frac{a'}{a} +
\frac{n'}{n}\Biggr) -\frac{b^2}{n^2}\,\Biggl[\frac{\dot{a}}{a}\,
\Biggl(\frac{\dot{a}}{a}-\frac{\dot{n}}{n}\Biggr) +
\frac{\ddot{a}}{a}\Biggr]\Biggr\} = \hat{\kappa}^2\,\hat{T}_{55}\,, 
\label{55}
\ea
where $\hat{\kappa}^2=8\pi \hat{G}=8\pi/M_5^3$ and the dots and primes
denote differentiation with respect to $t$ and $y$, respectively. 

The 5-dimensional energy-momentum tensor of the theory is obtained by the
variation of $\sqrt{-G^{(5)}}\,\hat{\cal L}_o$ with respect to the metric
and may be written in the form
\be
\hat{T}^M_{\,\,\,\,\,N}={\rm diag} (- \hat{\rho}, \hat{p}, \hat{p}, \hat{p},
\hat{T}^5_5)\,,
\ee
and, when combined with the equation for the conservation of the
energy-momentum tensor, $D_M \hat{T}^M_{\,\,\,\,\,\,N}=0$, leads to the
following relations 
\ba
\frac{d\hat{\rho}}{dt} + 3(\hat{\rho}+\hat{p})\,\frac{\dot{a}}{a} 
+ (\hat{\rho} + \hat{T}_5^5)\,\frac{\dot{b}}{b} &=& 0\,,
\label{zeroth} \\[5mm]
\Bigl(\hat{T}^5_5\Bigr)^{'} + \hat{T}^5_5\Biggl(\frac{n'}{n} +
3\frac{a'}{a}\Biggr)
+ \frac{n'}{n}\,\hat{\rho} -3 \frac{a'}{a}\,\hat{p} &=& 0\,.
\label{fifth}
\ea
In what follows, we adopt the brane-world scenario and, thus, we assume
that the usual matter content of the universe is confined on a brane which
may have zero or non-zero thickness. However, we make no such
assumption regarding the fifth component
$\hat{T}^5_5$ which can be smoothly distributed along the extra dimension
while non-vanishing $\hat{\rho}$ and $\hat{p}$ may exist in the bulk
due to a cosmological constant. 

In the framework of the above theory, we are looking for cosmological
solutions which admit compactification of the extra dimension and, at
the same time, respect the usual form of the 4-dimensional Friedmann
equation that arises from Einstein's equations in the limit
$y \rightarrow 0$. Since we are
interested in a compact extra dimension, we impose the condition 
that the scale factor of the fifth dimension depends neither on space
nor time, i.e. $b=b_0=const$. In that case, our task is greatly simplified
since the $(05)$-component (\ref{05}) of Einstein's equations can
be easily integrated to give the result
\be
n(t,y)=\lambda(t)\,\dot{a}(t,y)\,.
\label{soln}
\ee
Moreover, the $(00)$-component (\ref{00}), now, reduces to a second-order 
differential equation for $a(t,y)$ with respect to $y$ with the general
solution depending on the form of the energy density $\hat{\rho}$ of the
universe. The function $\lambda(t)$ is of equal importance as it leads,
not only to the determination of the lapse function $n(t,y)$, but also
to the 4-dimensional Friedmann equation at the origin, that is, on the
brane. Using the normalization
$n(t,y=0)=1$, the Hubble parameter can be expressed in terms of $\lambda(t)$
in the following way
\be
H^2 \equiv \biggl(\frac{\dot{a}_0}{a_0}\biggr)^2=
\frac{1}{\lambda^2(t) a_0^2(t)}\,,
\label{hubble}
\ee
where the subscript $0$ denotes quantities evaluated at $y=0$. As it will
be demonstrated shortly, the form of the function $\lambda(t)$, and
thus the restoration of the Friedmann equation at the origin, will strongly
depend on the expression of the fifth component of the energy-momentum
tensor $\hat{T}^5_5$.

It is important to remark at this point, that so far, we have considered
the constancy of the scale factor $b$ as an external condition.  That
is, it is not one which is obtained automatically from the solution of
Einstein's equations.  Fixing $b$, is equivalent to fixing the vev of the
dilaton in the dimensionally reduced theory.
This is clearly a physically reasonable assumption since 
a rolling dilaton implies the variation of gauge couplings and particle
masses and there are very strong
constraints against this \cite{co}.
Indeed one would expect that a dilaton potential is generated elsewhere
in the theory (buried in $\hat{\cal L}_o$) and provides for the necessary
additional condition which fixes $b$. As we shall see in the next section,
such a potential automatically gives rise to the $T_{55}$ component
advocated in \cite{kkop1}. 

\sect{Stabilization of the extra dimension}


Suppose for a moment that we consider the simple example of a 5D
Kaluza-Klein theory without matter. As such, we will have a solution for
any size of the compactified dimension, i.e. the
dilaton potential is flat.  Let us now suppose that 5D matter with a
given energy-momentum tensor is distributed in this space. 
Clearly, there is no guarantee that a constant radius will remain 
an integral of motion unless the energy-momentum of the matter satisfies
certain constraints. We next consider the conditions which must be
imposed on $\hat T^M_{\,\,\,\,N}$ in order to preserve the dilaton flatness. 

In some sense the answer to this question is 
similar to the following toy example. 
Imagine a massless scalar field $\phi(x)$ in $d$ dimensions which 
interacts with an external current $J(x)$. The condition 
\be
\int J(x)d^dx=0.
\label{toy}
\ee
ensures the stability of any vacuum configuration 
$\langle\phi(x)\rangle = v$ at the classical level.  
An analogous integral condition exists in the case of the 
Kaluza-Klein theory. The dilaton remains in equilibrium and 
the size of an extra dimension stays fixed if  the stress 
energy of the 5D matter satisfies the following constraint\footnote{This
constraint was also derived from a topological argument in \cite{E}.}:
\be
\int d^4x\,dy\,\sqrt{-G^{(5)}}\,(\hat{T}^\mu_\mu-2\hat{T}^5_5)=0.
\label{stable1}
\ee

To show how this condition actually emerges we work out an example of 
$\hat T^M_{\,\,\,\,N}$ created by a 5D scalar field, noting that the same 
arguments  can be extended for other types of matter.
We start by making a standard Kaluza-Klein decomposition of our diagonal
metric (\ref{metric}) in the following way
\be
G^{(5)}_{MN}= \left(\begin{tabular}{cc} $G^{(4)}_{\mu\nu}$ & $0$\\[1mm]
$0$ & $e^{2\gamma}$\end{tabular} \right)\,,
\label{kk}
\ee
where $G^{(4)}_{\mu\nu}$ denotes the usual 4-dimensional metric and
$e^{2\gamma(t)} \equiv b^2(t)$. Here, for the sake of our analysis, we
restore the time-dependence of the scale factor along the extra dimension.

We also consider the following 5-dimensional field theory of a single
scalar field minimally coupled to gravity
\be
S=- \int\, d^4x\,dy\,\sqrt{-G^{(5)}}\,\Bigg\{
\frac{R^{(5)}}{2\hat{\kappa}^2}+ \frac{1}{2}\,\partial_M \phi
\,\partial^M\phi + V(\phi)\Biggr\}\,.
\ee
where $\hat{\kappa}^2 = 8 \pi / M_5^3$.
Under the aforementioned decomposition (\ref{kk}), the above theory
can be written as
\be
S=- \int\, d^4x\,dy\,\sqrt{-G^{(4)}}\,e^\gamma\,
\Biggl\{\frac{1}{2\hat{\kappa}^2}\,
\Bigl[R^{(4)}(t,y) - {2}{e^{-\gamma}}\, D_\mu D^\mu e^\gamma\Bigr]
+ \frac{1}{2}\,\partial_M \phi\,\partial^M\phi + V(\phi) \Biggr\}
\ee
The above form of the theory is not particularly convenient for our
analysis: the ``kinetic" term for the dilaton field $\gamma$ is a
total derivative, and, thus, drops out from the action and, moreover,
there is a direct coupling between $\gamma$ and $R^{(4)}(t,y)$. Both
of these problems can be overcome by making the conformal transformation
$G^{(4)}_{\mu\nu} = e^{-\gamma} g_{\mu\nu}$. Then, the action functional
of the theory takes the final form
\be
S=- \int\, d^4x\,dy\,\sqrt{-g}\,\Biggl\{\frac{1}{2\hat{\kappa}^2}\,
\Bigl[R^{(4)}(t,y) +\frac{3}{2}\,\partial_\mu\gamma\,\partial^\mu\gamma\Bigr]
+ \frac{1}{2}\,\partial_\mu \phi\,\partial^\mu\phi 
+ \frac{1}{2}\,e^{-3\gamma}\,(\partial_5 \phi)^2
+ e^{-\gamma}V(\phi) \Biggr\}\,.
\label{action3}
\ee
Our next task is to derive the equation of motion for the dilaton field
$\gamma$ and find the condition that the components of the energy-momentum
tensor of matter need to satisfy in order to ensure the stability of the
extra dimension. The variation of the action (\ref{action3}) with respect
to $\gamma$ is given by the expression
\be
\delta S =\int dy\,\sqrt{-g}\left(
\frac{3}{2\hat{\kappa}^2}\,\,D_\mu D^\mu \gamma 
-\Bigl[-\frac{3}{2}\,e^{-3\gamma}\,(\partial_5 \phi)^2
- e^{-\gamma} V(\phi)\Bigr]\right)\delta\gamma\,,
\ee
from which we learn that the stability condition, analogous to 
(\ref{toy}) is
\be
\int dy\,\sqrt{-g}\Bigl[\frac{3}{2}\,e^{-3\gamma}\,(\partial_5 \phi)^2
+ e^{-\gamma} V(\phi)\Bigr]=0.
\label{intcond}
\ee

This constraint can be expressed in terms of the components
of the energy-momentum tensor generated by the 5-dimensional scalar field.
Before making the conformal transformation of the 4-dimensional metric, 
$\hat{T}_{MN}$ had the form
\be
\hat{T}_{MN}=\partial_M\phi\,\partial_N\phi
-G^{(5)}_{MN}\,
\Bigl[\frac{1}{2}\,\partial_P \phi\,\partial^P \phi + V(\phi)\Bigr]\,.
\ee

Then, we can easily find the following components
\ba
&~& \hat{T}^\mu_\mu=G^{(4)\,\mu\nu}\,\hat{T}_{\mu\nu}=
-\partial_\mu\phi\,\partial^\mu\phi-
2e^{-2\gamma}\,(\partial_5\phi)^2 -4 V(\phi)\,,\\[3mm]
&~& \hat{T}^5_5=e^{-2\gamma}\,\hat{T}_{55}=
-\frac{1}{2}\,\partial_\mu\phi\,\partial^\mu\phi+
\frac{1}{2}\,e^{-2\gamma}\,(\partial_5\phi)^2 - V(\phi)\,.
\ea
We can now rewrite the constraint equation (\ref{intcond}) preventing
a linear  source for the dilaton $\gamma$  in terms of $\hat{T}_{MN}$
which up to a multiplicative constant yields
\be
\int dy\,\sqrt{-g}\,\frac{e^{-\gamma}}{2}\,(\hat{T}^\mu_\mu-2\hat{T}^5_5)=0\,.
\label{stable2}
\ee
Note that, although $\hat{T}^\mu_\mu$ and $\hat{T}^5_5$ are not conformally
invariant, their combination $\hat{T}^\mu_\mu-2\hat{T}^5_5$ is, since the
difference does not depend explicitly on $G^{(4)}_{\mu\nu}$.
Therefore, the above result remains invariant under the conformal
transformation that we impose on the  4-dimensional part of the metric
tensor. 

We can easily demonstrate that the introduction of any type of matter
in the context of the 5D theory serves as a source term for the 4D gravitons.
More specifically, consider the linear term 
\be
\delta G^{(5)}_{MN}\,\hat T^{MN}=\delta G^{(4)}_{\mu\nu}\,\hat T^{\mu\nu} +
\delta (e^{2\gamma})\,\hat T^{55}
\ee
in {\em any} system. Using $G^{(4)}_{\mu\nu} = e^{-\gamma}
g_{\mu\nu}$, one can rewrite the above expression as
\be
\delta G^{(5)}_{MN}\,\hat T^{MN} = 
e^{-\gamma} \delta g_{\mu\nu}\,\hat T^{\mu\nu}
- \delta\gamma\,(\hat T^\mu_\mu-2\hat T^5_5)
\ee
Thus, we see that $\hat T_{\mu\nu}$ is a source for the graviton while the
combination $(\hat T^\mu_\mu-2\hat T^5_5)$
is the source for the dilaton or for any Brans-Dicke scalar field.
In the presence of a mechanism that stabilizes the extra dimension,
the source for the dilaton is identically zero. If we do not have
a stabilization mechanism but, nevertheless, we look for solutions
with a Newtonian limit, we still have to require the absence of the above
term. On the other hand, the presence of a scalar field in the theory
will give rise, not to a standard Einstein theory of gravity, but to
a scalar-tensor theory with all its consequences.



All of these considerations help to elucidate the required form 
for $\hat T^M_{\,\,\,\,N}$ in the 5-dimensio\-nal space which allows us to keep 
the size of the extra dimension fixed. 
Eq. (\ref{stable1}) is the constraint which in fact must be satisfied.
This constraint can, for example,  be easily specialized to the
cases already considered  in previous publications. Following \cite{BDL},
one may choose two  branes with opposite matter densities and vanishing
$\hat T_5^5$. In this case  the stability condition  (\ref{stable1}) is
trivially satisfied (since the $\hat T^\mu_\mu$'s of each brane cancel
each other).  As another illuminating example, we consider the thin-wall
cosmological solution that was derived in Ref. \cite{kkop1} and describes
the 5-dimensional space-time around an infinitely thin brane-universe
located at $y=0$. The ordinary matter density is confined to the
brane and no other form of energy -- apart from a smooth distribution of
$\hat{T}^5_5$ -- exists in the bulk. The energy-momentum tensor of the
brane-universe, in this case, has the form
\be
\hat{T}^M_{\,\,\,\,N}={\rm diag} \Bigl(\frac{\delta(y)}{b_0}(-\rho, p, p, p),
\hat T^5_5\Bigr)\,.
\label{tmn}
\ee
The expression for $\hat{T}^5_5$ in the bulk follows from
the assumption that stable extrema of $n(y)$ and $a(y)$ 
exist outside the brane, at $y=\pm |y_{min}|$, and that
the Friedmann equation has the usual form at the origin.
The resulting energy-momentum tensor component is
\be
\hat{T}_5^5=\frac{a_0^3(t)}{2 n(t,y) a^3(t,y)}\,\frac{(-\rho+3p)}
{2b_0|y_{min}|} + {\cal O}(\rho^2)\,.
\label{ansatz}
\ee
Note that, in our expansion in powers of $\rho$, the prefactor
$(a_0^3/n a^3) = 1+ {\cal O}(\rho^2)$, so that
$\hat T^5_5=(-\rho+3p)/(2b_0|y_{min}|) + {\cal O}(\rho^2)$.

Going back to eq. (\ref{tmn}), the 4-dimensional trace of the energy-momentum
tensor is simply given by the expression
\be
\hat{T}^\mu_\mu =\frac{\delta(y)}{b_0}\,(-\rho + 3p)\,.
\ee
Taking into account the fact that the 
4-dimensional part of the energy-momentum
tensor is localized on the brane while $\hat{T}^5_5$ is not and performing
the integration in (\ref{stable2}) with respect to 
$y \in (-|y_{min}|, |y_{min}|)$, we can easily demonstrate that the
desirable form of $\hat T_5^5$ satisfies our general constraint (\ref{stable1}).

The general condition on the stability of the radius 
(\ref{stable1}) represents a highly nontrivial fine-tuning 
problem for the matter fields in 5D space  {\em if} one
insists on a flat potential for the dilaton and on the compactification 
at an arbitrary value of $\gamma$. Moreover, the dilaton in
this scenario will remain massless, which constitutes another
serious phenomenological problem as noted above. Therefore, we must
consider a more realistic situation with an 
additional physical input, namely an explicit 
mechanism which gives the mass to the dilaton and 
stabilizes it around a specific local minimum.

It is tantalizing to write this potential in the simplest possible 
form, 
\be
V(G^{(5)}_{55}) = \alpha( G^{(5)}_{55} - b_0^2 )^2, 
\label{naive}
\ee
so that $G^{(5)}_{55}=e^{2\gamma}=b_0^2$ represents an equilibrium point.
Since this form explicitly breaks general covariance, it must be only {\em
effective}, realized by some bulk mechanism,  unspecified at this point. 
Any uniform deviation of $G^{(5)}_{55}$ from $b_0^2$ induces a uniform 
value for $\hat T_5^5$ in the bulk, given by the derivative of $V$ with
respect to $G^{(5)}_{55}$. If a brane with the non-vanishing matter energy
density is included, the equilibrium  position of
$G^{(5)}_{55}$ changes. We next show that the  value of $\hat T_5^5$,
generated this way, does satisfy eq. (\ref{ansatz}) independent of
the parameters which characterize the stiffness of the dilaton potential.

To have a workable framework which preserves general covariance we 
choose an auxiliary bulk scalar $\chi$ with the following
unusual action
\be
S=-\int d^4x\int_{-|y_{min}|}^{|y_{min}|}\sqrt{-G^{(5)}}
dy\left(\partial_M \chi \partial^M \chi - c_1^2\right)^2
\label{chi}
\ee
in the given background configuration $\chi(y) = c_2|y|$. This background
configuration creates the potential of the form (\ref{naive}) if we
identify $b_0 = c_2/c_1$. The dilaton chooses its preferred point
$\gamma_0$ and  it is easy to check that
$G^{(5)}_{55}=e^{2\gamma_0}=b_0^2=c_2^2/c_1^2$ together  with  $\chi(y) =
c_2|y|$ satisfy all of the equations of motions.  At the same time, at the
equilibrium position, all the components of
$\hat T^M_{\,\,\,\,N}$ which correspond to the action (\ref{chi}) remain
identically zero. We also note here that such a background configuration 
for $\chi$ is consistent with compactification and it can be
further checked that with the choice of (\ref{chi}),
the cusps in $\chi(y)$ do not require the existence of 
delta-functional sources.

When a brane with matter content is introduced at
$y=0$, the equilibrium position for $\gamma$ changes. This change can be 
read off from the interaction of $\gamma$ with the trace of the 
brane energy-momentum tensor, eq. (\ref{stable2}), and the effective potential 
generated by (\ref{chi}). Keeping only the term linear in
this deviation, $\gamma - \gamma_0$, in front of the small 
perturbation created by $\hat T_\mu^\mu$ of the brane, we write down 
an effective potential for the dilaton of the form
\be
V_{eff}(\gamma)= \frac{8}{e^{5\gamma_0}}\,|y_{min}|\,c_2^4 (\gamma-\gamma_0)^2 - 
\frac{1}{2e^{2\gamma_0}}\,(-\rho+3p) (\gamma-\gamma_0),
\ee
from which we see that the new equilibrium point is 
given by 
\be
(\gamma-\gamma_0) = {e^{3\gamma_0}\over c_2^4} {(-\rho+3p) \over 32\,|y_{min}| }.
\label{shift}
\ee
This deviation produces a non-vanishing (55)-component of the 
energy-momen\-tum tensor in the bulk (from the variation of \ref{chi}),
exactly of the form discussed above. Indeed, neglecting all terms on the 
order of
$\rho^2$, we have the following value for $\hat T_5^5$
\be
\hat T_5^5 = 8 c_2^4 e^{-4\gamma_0}(\gamma -\gamma_0) =
{-\rho+3p \over 4 b_0 |y_{min}| },
\label{reaction}
\ee
whereas the (00) and $(ii)$ components of $\hat T^M_{\,\,\,\,N}$ in the bulk
remain zero to this accuracy. The same arguments hold for any mechanism
which would render an effective potential in the form (\ref{naive}).

Equation (\ref{reaction}), is exactly of the same 
value needed for the restoration of the Friedmann equation
for the brane scale factor, and
allows for an important interpretation of $\hat T_5^5$. Indeed,
we saw that this component arises as the backreaction of the 
dilaton fixing potential on the presence of the energy density on 
the brane. This confirms a hypothesis put forward in our previous 
work \cite{kkop1}. At the same time it removes the necessity of the 
fine tuning (\ref{stable2}) required for the case of the flat potential.
It turns out that the only requirements which should be 
imposed on $\alpha$ in (\ref{naive}) or $c_2^2$ in (\ref{chi})-(\ref{reaction})
is the condition $\gamma-\gamma_0\ll\gamma_0$ which justifies 
our linearized approach. 
We see that the details of the dilaton 
fixing potential are irrelevant to linear order in $\rho$ and affect
only the sub-leading, $\rho^2$-proportional terms. We note that mechanisms
for generating a dilaton potential from bulk fields was discussed
in \cite{GW}. 

Recent work on the stabilization of the dilaton in the context of
the  Randall-Sundrum model showed a similar shift in the dilaton
expectation value \cite{randall3}. Indeed, the shift found there and the
resulting change in the derivative of the dilaton potential at the
shifted point acts as a source to $\hat G_{55}$ giving an identical
expression for $\hat T_{55}$ described here and in \cite{kkop1}.

\sect{Solutions with cosmological constant and compactification}

In this section, we adopt the scenario of an infinitely thin brane-universe
located at $y=0$ with the ordinary energy density and pressure being
localized on the brane. We further assume that the total energy density
of the brane, $\rho_{br}$, is the sum of the ordinary matter density,
$\rho$, and a brane cosmological constant, $\Lambda_{br}$. On the other
hand, $\hat{T}^5_5$ is smoothly distributed along the extra dimension
while the bulk is dominated by the energy density coming from a bulk
cosmological constant, $\Lambda_B$. Here, we are going to investigate
the existence of cosmological solutions that admit compactification
of the extra dimension while leading, once again, to the standard
cosmological expansion of the brane scale factor. We will consider
three different cases corresponding to negative, positive and zero
bulk cosmological constant. 
The sign of $\Lambda_{br}$ will be
determined by demanding the cancellation of the effective cosmological 
constant in the Friedmann equation.
Related brane solutions with both brane and bulk cosmological constant
were considered in \cite{NK,BDEL}.

\subsection{Negative cosmological constant in the bulk}

We start with the case of a negative bulk cosmological constant,
$\Lambda_{B}<0$, and we write the energy-momentum tensor in the bulk 
in the following form 
\be
\hat{T}^A_B=\Bigl(-\rho_B, p_B, p_B, p_B, \hat{T}^5_5 \Bigr)\,,
\ee
where $\rho_B=\Lambda_B<0$ and the other components are, for the
moment, arbitrary. For a smooth, constant distribution of energy in the
bulk, the solution for the scale factor $a(t,y)$, outside the brane, will
follow from the (00)-component of Einstein's equations, which
takes the simple form
\be
x''-A^2x-B^2=0\,,
\ee
where $x \equiv a^2$ and
\be
A^2=\frac{2 \hat{\kappa}^2}{3}\,b^2 |\Lambda_B|\,\,, \qquad
B^2(t) =\frac{2b^2}{\lambda^2(t)}\,.
\ee

The general solution of the above differential equation has the form
\be
a^2(t,y)=d_1(t)\,\cosh(A|y|) + d_2(t)\,\sinh(A|y|) -\frac{B^2(t)}{A^2}\,.
\label{general2a}
\ee
The boundary conditions on the brane will determine the two unknown
functions of time, $d_1(t)$ and $d_2(t)$. Defining 
$a^2(t,y=0) \equiv a^2_0(t)$, we can write
\be
d_1(t)=a_0^2(t) +\frac{B^2(t)}{A^2}
\label{sold1}
\ee
while the {\it jump} in the first derivative of $a(t,y)$ across the origin,
which now takes the form
\be
\frac{[a']}{a_0\,b}=-\frac{\hat{\kappa}^2}{3}\,(\rho+\Lambda_{br})\,,
\ee
leads to the following expression for the second function of time
\be
d_2(t)=-\frac{\hat{\kappa}^2}{3}\,\frac{a_0^2\,b}{A}\,(\rho+\Lambda_{br})\,.
\label{sold2}
\ee

When the general solution (\ref{general2a}) is substituted in the
$(ii)$-component of Einstein's equations, we get the anticipated result
$p_B=|\Lambda_B|=-\rho_B$ while, the same procedure with the
(55)-component, leads to the constraint
\be
\frac{d}{dt}\,\biggl[\,\frac{A^2}{4}\,(d_2^2-d_1^2) +
\frac{b^4}{A^2 \lambda^4}\,\biggr]=\frac{2\hat{\kappa}^2}{3}\,b^2 a^3
\dot{a}\,(\hat{T}^5_5-|\Lambda_B|)\,.
\label{con55}
\ee
The above constraint will allow us to determine the form
of $\hat{T}^5_5$ once the expression for $\lambda(t)$ is fixed
by the Friedmann equation.

Next, we come to the question of the compactification. Even in the
absence of a second brane, the compactification of the extra dimension
could easily take place as long as the solution (\ref{general2a}) has a
stable extremum at some point $y=|y_{min}|$. Then, by 
identifying the points $y=\pm |y_{min}|$, the extra dimension would
be compactified with its size being $2b\,|y_{min}|$. From the
vanishing of $a'(t,y)$, we find that
\be
\tanh(A\,|y_{min}|) = -\frac{d_2(t)}{d_1(t)} \equiv c_0.
\label{minimum2a}
\ee
The above ratio, as indicated, must be constant in order for the
extremum to be stable in time. Then, by using the above relation and
eqs. (\ref{sold1}) and (\ref{sold2}), the Friedmann equation 
immediately follows
\be
\biggl(\frac{\dot{a}_0}{a_0}\biggr)^2=\frac{1}{\lambda^2 a_0^2}=
\frac{\hat{\kappa}^2}{3}\,|\Lambda_B| \,\biggl\{-1 +\frac{\hat{\kappa}^2}{c_0}\,
\frac{(\rho+\Lambda_{br})}{\sqrt{6 \hat{\kappa}^2 |\Lambda_B|}}\biggr\} \equiv
\frac{\kappa^2 \rho}{3}\,.
\label{fried}
\ee
The last equality in the above equation defines the 4D Planck constant and
holds only under the constraint
\be
-1 + \frac{\hat{\kappa}^2}{c_0}\,
\frac{\Lambda_{br}}{\sqrt{6 \hat{\kappa}^2 |\Lambda_B|}}=0
\label{con1}
\ee
and the redefinition
\be
\kappa^2= \frac{\hat{\kappa}^2}{c_0}\,
\sqrt{\frac{\hat{\kappa}^2 |\Lambda_B|}{6}}\,.
\label{kappa4}
\ee

These expressions can also be used to define the 4D cosmological constant
on the brane. From eq. (\ref{fried}), we can write
\ba
\Lambda_{eff} & = & \frac{\hat{\kappa}^2}{\kappa^2} |\Lambda_B|
\,\biggl\{-1 +\frac{\hat{\kappa}^2}{c_0}\,
\frac{\Lambda_{br}}{\sqrt{6 \hat{\kappa}^2 |\Lambda_B|}}\biggr\}
\nonumber \\
& = & \Lambda_{br} - c_0{\sqrt{{6 |\Lambda_B| \over \hat{\kappa}^2 }}}
\ea
In the limit that $A|y_{min}|$ is large, $c_0\simeq 1$ and 
$\Lambda_{eff} = \Lambda_{br} - {\sqrt{{6 |\Lambda_B| \over \hat{\kappa}^2
}}}$. In the other limiting case, when $A|y_{min}|$ is small,
we have $\Lambda_{eff} = \Lambda_{br} - (2 b\,|y_{min}|) \Lambda_B$.

{}From its definition (\ref{minimum2a}), since $A|y_{min}|$ is positive, it
is easy to see that
$c_0$ is always positive independently of the exact values of
$|\Lambda_B|$ and
$|y_{min}|$ and, thus, the Friedmann equation will always have the correct
sign. Then, in order for the constraint (\ref{con1}) to be satisfied, 
we must necessarily have $\Lambda_{br}>0$. The above two equations can be
rewritten in a simple form as
\be
|\Lambda_B|=\frac{\hat{\kappa}^2}{6\,c_0^2}\,\Lambda_{br}^2\,\,, \qquad
\kappa^2 =\frac{\hat{\kappa}^4}{6\,c_0^2}\,\Lambda_{br}\,.
\label{redef}
\ee

Finally, we need to determine the form of $\hat{T}^5_5$ in the bulk. By
using the constraint (\ref{con55}) and the relations (\ref{redef}), we find
that
\be
\hat{T}^5_5=|\Lambda_B| + \frac{\hat{\kappa}^2\,a_0^3}{12\,n(t,y)\,a^3(t,y)}\,
\bigg\{-\rho\,(\rho + 3p) + \Bigl(1-\frac{1}{c_0^2}\Bigr)\,
\Lambda_{br}\,(2 \Lambda_{br} +\rho -3p) \Biggr\}
\label{552a}
\ee
A number of comments should be made at this point. In the limit
$|y_{min}| \rightarrow \infty$, the extra dimension is non-compact and 
$c_0=1$. In that case, the last term in the above expression becomes zero
and, modulo a small correction of ${\cal O}(\rho^2)$, $\hat{T}^5_5$
assumes the form of another pressure-like component of the energy-momentum
tensor, equal to $|\Lambda_B|$. This is the case considered in Ref.
\cite{BDEL}. In order to compactify, their solution requires the
introduction of a second brane along with an unavoidable
correlation between the energy densities on the two branes.  In
particular, the absence of $\hat T^5_5$, again implies the presence of a
negative energy brane.  
For every other value of $c_0$, our results describe 5-dimensional
cosmological solutions that respect the 4-dimensional Friedmann equation
and allow for the compactification of the fifth dimension via the
existence of an extremum of $a(t,y)$ at some finite point $y=|y_{min}|$. 
As we have seen
in the previous section, the deviation of the fifth component of the
energy-momentum tensor from the strict limits of a pressure-like
component, is crucial for the stabilization of the extra dimension.

Let us, finally, note that the compactification of the extra dimension
via the extremum of the scale factor not only eliminates the need
for the introduction of a second brane but also resolves a major problem
that arises in the framework of the two-brane models. For
$\Lambda_B<0$, it has been shown
\cite{Berkl,Cline} that, in order to reproduce the correct sign in
the Friedmann equation, one of the two branes must have an energy
density which is negative, $\rho<0$. The
need for this unnatural assumption ceases to exist in the
context of our analysis. Despite the fact that we start with the same
assumption of a negative cosmological constant in the bulk, in our case,
the correct sign in the Friedmann equation is always maintained.
Two- and multi-brane solutions in this context
will be presented elsewhere.

\subsection{Positive cosmological constant in the bulk}

In this case, we assume that $\rho_B=\Lambda_B>0$. The analysis closely
follows the one of the previous subsection, however, the opposite sign
of $\Lambda_B$ modifies some of our results and, consequently, our final
conclusions. We start again with the general solution for the scale
factor $a(t,y)$ outside the brane, which, now, has the form
\be
a^2(t,y)=d_1(t)\,\cos(A|y|) + d_2(t)\,\sin(A|y|) +\frac{B^2}{A^2}\,.
\label{general2b}
\ee
The first coefficient $d_1(t)$ is given by eq. (\ref{sold1}) with $A^2$
being replaced by $-A^2$ while the second one, $d_2(t)$, is still
given by eq. (\ref{sold2}).
Once again, it is easy to see that the above solution satisfies the
$(ii)$-component of Einstein's equations provided that $p_B=-\Lambda_B=-\rho_B$
while, the (55)-component, leads to the constraint
\be
\frac{d}{dt}\,\biggl[\,\frac{A^2}{4}\,(d_2^2+d_1^2) -
\frac{b^4}{A^2 \lambda^4}\,\biggr]=\frac{2\hat{\kappa}^2}{3}\,b^2 a^3
\dot{a}\,(\hat{T}^5_5+\Lambda_B)\,.
\label{con552}
\ee

We still need a stable extremum for the compactification of the extra
dimension and this extremum, now, takes place at 
\be
\tan(A\,|y_{min}|) = \frac{d_2(t)}{d_1(t)} \equiv c_0\,.
\label{minimum2b}
\ee
By using the above relation and the expressions for $d_i(t)$, the Friedmann
equation can, now, be written as
\be
\biggl(\frac{\dot{a}_0}{a_0}\biggr)^2=
\frac{\hat{\kappa}^2}{3}\,\Lambda_B \,\biggl\{ 1 +\frac{\hat{\kappa}^2}{c_0}\,
\frac{(\rho+\Lambda_{br})}{\sqrt{6 \hat{\kappa}^2 \Lambda_B}}\biggr\}\,.
\ee
The standard form of the above equation is immediately restored under the
constraint
\be
1 + \frac{\hat{\kappa}^2}{c_0}\,
\frac{\Lambda_{br}}{\sqrt{6 \hat{\kappa}^2 \Lambda_B}}=0\,,
\label{con2}
\ee
while the redefinition of the 4-dimensional Newton's constant in terms of
the 5-dimensional one is still given by eq. (\ref{kappa4}). From its
definition, eq. (\ref{minimum2b}), we conclude that $c_0$ is not always
positive. Thus, in order to have the correct sign in the Friedmann equation, 
we have to impose the following constraint on the possible values of
$|y_{min}|$ 
\be
n\,\pi < A\,|y_{min}| < (n+\frac{1}{2})\,\pi\,, \qquad n \in Z\,.
\ee
Then, the constraint (\ref{con2}) can be satisfied only in the case 
where $\Lambda_{br}< 0$. Note that the cosmological constants in the bulk
and on the brane must always have opposite signs in order to ensure the 
cancellation of their contributions to the energy density of the universe.
However, their sign does not enter the Friedmann equation neither
imposes any unnatural constraints on the energy density of our
brane-universe.

Let us note that, contrary to the case where $\Lambda_B<0$, here, there are
no solutions that would correspond to a non-compact extra dimension.
Every cosmological solution, described by eqs. (\ref{general2b}) and
(\ref{minimum2b}), that respects the Friedmann equation at the origin
has a compact extra dimension with finite size $2b \,|y_{min}|$. 
This is also obvious from the expression for the fifth component of the
energy-momentum tensor in the bulk, which from the constraint (\ref{con552})
is found to be 
\be
\hat{T}^5_5=-\Lambda_B + \frac{\hat{\kappa}^2\,a_0^3}{12\,n(t,y)\,a^3(t,y)}\,
\bigg\{-\rho\,(\rho + 3p) + \Bigl(1+\frac{1}{c_0^2}\Bigr)\,
\Lambda_{br}\,(2 \Lambda_{br} +\rho -3p) \Biggr\}\,.
\label{552b}
\ee
Note that the form of $\hat{T}^5_5$ is always different from the one of
a pressure-like component and there is no real value of $c_0$
that would eliminate the last term. Thus, in order to stabilize the extra,
compact dimension, a non-trivial $\hat T^5_5$ must be generated.

\subsection{Zero cosmological constant in the bulk}

Finally, we consider the case where $\Lambda_B=0$ while keeping a non-zero
cosmological constant on the brane. In the
bulk, the solution is exactly the same as in the case
studied in the thin-wall approximation of our previous work \cite{kkop1}.
The general solution for $a(t,y)$ can be written as
\be
a^2(t,y)=a_0^2(t) + c(t)\,|y| + \frac{b^2}{\lambda^2}\,y^2\,,
\label{general2c}
\ee
with
\be
c(t)=-\frac{\hat{\kappa}^2}{3}\,a_0^2\,b\,(\rho+\Lambda_{br})\,.
\ee
For $\rho+\Lambda_{br}>0$, the above solution describes a spatial
scale factor that decreases as we move away from the brane. However,
the rate of decrease is much smaller than the one given by the
warp factor in the Randall-Sundrum model \cite{RS}.

The compactification of the extra dimension is made by
identifying the two extrema which, now, take place at the value
\be
|y_{min}|=-\frac{c(t)\,\lambda^2(t)}{2b^2}\,,
\ee
and, by using the above result for $c(t)$, the Friedmann equation may
be written as
\be
\biggl(\frac{\dot{a}_0}{a_0}\biggr)^2=
\frac{\hat{\kappa}^2}{3}\,\frac{(\rho+\Lambda_{br})}{2b\,|y_{min}|}=
\frac{\kappa^2}{3}\,(\rho+\Lambda_{br})\,.
\ee
Due to the absence of a bulk cosmological constant that would cancel the 
contribution of $\Lambda_{br}$ to the total energy of the universe,
the term linear in $\rho$ will become dominant only in the case
$\Lambda_{br} \ll \rho$. In the opposite case, as anticipated, the presence
of $\Lambda_{br}$ will significantly modify the expansion rate of the scale
factor at the origin. 

\sect{Thick-wall solution with compactification}

To complete the class of solutions we have been discussing, we generalize
our thin-wall solution found in \cite{kkop1} to one of finite thickness. 
We assume that our brane-universe has a non-zero thickness
$2\Delta b$ with the normal matter content being distributed throughout the
brane. Outside the brane, both the energy density and pressure are
zero, which leads, after integrating the (00)-component of Einstein's
equations, to the following general solution for the spatial scale factor
\be
a^2(t,y)=a^2_\Delta(t) + C(t)\,(|y|-\Delta) + 
\frac{b^2}{\lambda^2}\,(y^2-\Delta^2)\,,
\label{general1}
\ee
where $a_\Delta(t)$ is the scale factor evaluated at $y=\Delta$ and
$C(t)$ an unknown function of time which needs to be determined.
For the compactification of the extra dimension, we require an
extremum in $a(t,y)$ outside the brane, so, we impose the
condition $a'(t,y_{min})=0$ on the general solution (\ref{general1}).
That leads to the result
\be
|y_{min}|=-\frac{C(t) \lambda^2}{2b^2}\,.
\label{minimum1}
\ee
In Ref. \cite{kkop1}, we showed that, in the thin-wall approximation, the
existence of a stable minimum for $a(t,y)$ outside the brane and, thus,
the compactification of the extra dimension is equivalent to the
restoration of the Friedmann equation at the origin. Here, we need
to impose that this equation takes the usual form, i.e. 
\be
\biggl(\frac{\dot{a}_0}{a_0}\biggr)^2=\frac{1}{\lambda^2 a_0^2}=
\frac{\kappa^2 \rho(t,0)}{3}
\label{friedmann}
\ee
where $\rho(t,0)$ is the usual 4-dimensional energy density defined at the
origin, and
\be
\kappa^2=\frac{\hat{\kappa}^2}{2b\,|y_{min}|}\,.
\ee
Then, by combining eqs. (\ref{minimum1}) and (\ref{friedmann}), we obtain
\be
C(t)=-\frac{\hat{\kappa}^2 \rho(t,0)\,a_0^2\,b}{3}\,.
\label{c}
\ee

It is easy to see that the solution (\ref{general1}) trivially satisfies
the $(ii)$-component of Einstein's equations while, when substituted in
the (55)-component, it leads to the constraint
\be
\frac{d}{dt}\,\biggl[\,\frac{b^2}{\lambda^4}\,(\Delta-|y_{min}|)^2 -
\frac{a_\Delta^2}{\lambda^2}\,\biggr] =\frac{2\hat{\kappa}^2}{3}\, 
a^3 \dot{a}\,\hat{T}^5_5\,.
\label{55a}
\ee
The above constraint, with the expression for $\lambda$ given by eq.
(\ref{friedmann}), is satisfied only if the fifth component of the 
energy-momentum tensor, in the bulk, has the form
\be
\hat{T}^{5}_{5\,(B)}(t,y)=\frac{1}{2n a^3}\,\bigg\{\frac{a_\Delta a_0}
{2b |y_{min}|}\,[a_\Delta\,(\rho + 3p) -2\rho\,a_0\,n_\Delta] -
\frac{\hat{\kappa}^2 a_0^3}{6}\,\frac{(\Delta-|y_{min}|)^2}{y_{min}^2}\,
\rho\,(\rho+3p) \biggr\}\,,
\label{55out}
\ee
where again $\rho$ and $p$ are defined at the origin. In the limit
$\Delta \rightarrow 0$, the above result reduces to what we had for the
thin-wall solution~\cite{kkop1}. We emphasize once more, that despite the
peculiar nature of the expression in (\ref{55out}), so long as the
dilaton picks up a stable expectation value, this is the form that
$T_{55}$ will assume. 

To complete the thick wall solution, we must derive the expression for 
$\hat{T}^{5}_{5}(t,y)$ inside the brane.  This derivation is given in the
Appendix, the result is 
\be
\hat{T}^5_{5\,(br)}(t,y)= \hat{T}^5_{5\,(B)}(t,y)
+ \hat{\rho}\,\biggl[\frac{n(t,\Delta)}{n(t,y)}-1 \biggr]\,,
\label{55in}
\ee
where $\hat{T}^5_{5\,(B)}(t,y)$ is given by eq. (\ref{55out}).
We can easily show that the above expression satisfies the fifth
component of the equation for the conservation of energy (\ref{fifth})
and that, in the limit $y \rightarrow \Delta$, the second term at the
r.h.s. of eq. (\ref{55in}) vanishes, thus, ensuring the continuity
of $\hat{T}_5^5$ across the boundary.

There is one remaining function we must determine, and that is the scale
factor on the boundary of the brane, $a_\Delta(t)$.  Again, we leave the
derivation to Appendix, and simply display the result here
\be
a_\Delta(t)=a_0(t)\,\Bigl[\,1-\frac{\hat{\kappa}^2}{12}\,
\rho(t,0)\,\Delta b\,\Bigr]\,.
\label{adin}
\ee
The above result reveals the fact that the scale factor $a(t,y)$ decreases
as we move away from the origin. Nevertheless, the rate of decrease being
simply proportional to the energy density on the brane is much smaller than
in the case of the Randall-Sundrum thin-wall solution \cite{RS} where the
rate of decrease depends exponentially on the bulk cosmological constant.
Due to the continuity condition on $a'(t,y)$ at the boundary, the scale
factor will also decrease mildly outside the brane until it reaches its
unique minimum at $y=\pm |y_{min}|$. The mild decrease of the scale factor
makes necessary the compactification of the extra dimension which is realized
through the identification of the two stable extrema $\pm |y_{min}|$.

In this section, we have demonstrated that the positive features exhibited
by the thin-wall solution found in \cite{kkop1} also characterize its
thick-wall analog. The cosmological
solutions that emerge respect the 4-dimensional Friedmann equation on
the brane and the compactification of the extra dimension takes place
without the need for a second brane. In both instances, the existence of
a non-vanishing $\hat T^5_5$ plays an important role as it results
from the stabilization the extra dimension and helps restore standard
cosmology on the brane.

\section{Conclusions}

While it may seem that standard FRW cosmology should arise trivially on a
homogeneous 3-brane embedded in a higher dimensional space time, it has become
clear that this result is highly dependent on the generation of a dilaton
potential which fixes the dilaton expectation value and hence the size of the
extra dimension.  In the absence of a dilaton potential, there are two serious
problems which must be overcome.  Aside from the non-compact solution of Ref.
\cite{RS2},  compactification of the extra-dimension in solutions for 
3-branes with matter require the presence of a second brane on which the
matter is constrained by Einstein's equations to take the form $\rho_2 =
-\rho_1 + O(\rho^2)$.  That is,  we are forced to tolerate negative energy
densities on the second brane.  This is most assuredly unphysical.
However, even if we accept this solution which includes the negative-brane, the
solutions to Einstein's equations yield a cosmological expansion for which 
the Hubble parameter is proportional to the energy density.  BBN and post
BBN-cosmology can not be reconciled with this expansion law. 
We stress that we distinguish here the difference between a brane with 
negative tension and one with negative matter density. While a negative tension
brane may naturally appear in string theory, we are not aware of any formulation 
which allows for a negative matter density $-\rho$. 

In our previous work, we proposed that both of the aforementioned problems were
tied to explicit assumptions made on the form of the energy-momentum tensor.
Namely, that 
\be
\hat{T}^M_{\,\,\,\,N}=\frac{\delta(y)}{b_0}{\rm diag} \Bigl(-\rho, p, p, p,
0\Bigr)\,.
\label{tmn0}
\ee
i.e., that the (55)-component of the energy-momentum tensor vanished. 
In \cite{kkop1}, we showed that by allowing $\hat T_{55}$ to differ from zero,
we
could find a solution for the 3-space scale factor $a(t,y)$, in the bulk, which
has a local minimum at some co-ordinate value $y_{min}$, at which point we
can compactify the extra dimension {\em without} the introduction of a second
brane and in particular without a negative energy brane of any kind. 
Secondly, we derived the form of $\hat T^5_{5}$, in order to recover a normal
Friedmann expansion on the 3-brane.

Here we have justified the choice for $\hat T^5_{5}$, and showed that it is the
natural result of the back-reaction of a stabilized dilaton to the 3-brane with
matter. In particular we have showed that the dilaton couples to the
combination $\hat T^\mu_\mu - 2 \hat T^5_5$, and {\em any} stable dilaton
configuration requires this combination of energy-momentum components to
vanish when averaged over the extra dimension. Furthermore, when a dilaton
potential, which fixes the dilaton vev is included in the derivation of the
solution, matter on the 3-brane will induce a shift in the dilaton vev
in such
a way so as to exactly guarantee the vanishing of $\hat T^\mu_\mu - 2 \hat
T^5_5$, and produce the needed (55)-component derived in \cite{kkop1}. 

We have, in addition, supplied generalization of our 3-brane solutions to
include a cosmological constant both in the bulk and on the brane. We
have considered the cases of negative, positive and zero bulk cosmological
constant and demonstrated that a solution with a stable, compact extra
dimension, which respects the standard 4D Friedmann equation, arises in each
case. Moreover, the correct sign in the Friedmann equation is always
maintained for every sign of $\Lambda_B$ as long as the sign of $\Lambda_{br}$
is exactly opposite. In this case, the contributions of the two
cosmological constants to the energy density of the universe can be
cancelled leaving a term linear in $\rho$ to govern the expansion of the
brane scale factor. Finally, our single-brane thin-wall solution
derived in \cite{kkop1} was extended in the case of a brane-universe
with a finite thickness. The resulting cosmological solution allows
for the compactification of the extra dimension through the identification
of the two stable
minima of the scale factor outside the brane and the 4D Friedmann equation
is once again recovered. In all of the above cases, the existence of
a non-trivial fifth component of the energy-momentum tensor is necessary
for the stabilization of the size of the extra dimension.
\paragraph{}

{\bf Acknowledgments} We would like to thank G. Ross, V. Rubakov, M. Shifman
and M. Voloshin for useful discussions. This work was supported in part by
the Department of Energy
under Grant No.\ DE-FG-02-94-ER-40823 at the University of Minnesota.
The work of I.I.K. is  supported in part by PPARC rolling grant
PPA/G/O/1998/00567, the EC TMR grant FMRX-CT-96-0090 and  by the INTAS
grant RFBR - 950567.

\def\theequation{A.\arabic{equation}}
\setcounter{equation}{0}
\vskip0.8cm
\noindent
\centerline{\Large \bf Appendix}
\vskip0.4cm
\noindent

In this Appendix, we show the details of the derivation of the energy-momentum
tensor inside the brane and subsequently generalize this solution for an
arbitrary equation of state.  For reasons, that will become obvious, we may
rewrite the expression for the scale factor outside the brane (\ref{general1})
as follows
\be
a^2(t,y)=\frac{b^2}{\lambda^2(t)}\,[\,|y|+C_{out}(t)]^2 -
\frac{\lambda^2(t) E_{out}(t)}{2b^2}\,,
\label{solout}
\ee
where
\be
C_{out}(t)= \frac{\lambda^2}{2b^2}\,C(t)=-|y_{min}|\,, \qquad
E_{out}(t)=2b^2\,\biggl[\,\frac{b^2}{\lambda^4}\,(\Delta-|y_{min}|)^2 -
\frac{a_\Delta^2}{\lambda^2}\,\biggr]\,.
\label{eout}
\ee
\par
To obtain the inside solution for the scale factor $a(t,y)$,
we will assume, at first, that the pressure on the brane is zero,
so our results would apply for a matter-dominated universe. Then, the
zeroth-component of the equation for the conservation of energy (\ref{zeroth})
gives $\hat{\rho}=\hat{\rho}_0/a^3$. Integrating the (00)-component
of Einstein's equations, the general solution for the scale factor takes
the implicit form
\ba
&~& \hspace*{-1.5cm}\frac{2A^2}{B^3(t)}\,\log\Biggl(\frac{2}{B(t)}\,
\Bigl[B^2(t) a(t,y)-A^2\Bigr] +
2 \sqrt{E_{in}(t) + B^2(t) a^2(t,y) -2 A^2 a(t,y)} \Biggr) \nonumber\\[4mm]
&~& \hspace*{1.5cm} +\, \frac{2}{B^2(t)}\, \sqrt{E_{in}(t) +
B^2(t) a^2(t,y) -2 A^2 a(t,y)}=
\pm \sqrt{2}\,[|y|+ C_{in}(t)]\,,
\label{sola}
\ea
where
\be
A^2= \frac{2b^2 \hat{\kappa}^2 \hat{\rho}_0}{3}\,.
\qquad B^2(t)=\frac{2b^2}{\lambda^2(t)}\,.
\ee
This is the same expression as the one that we derived in the thick-wall
approximation \cite{kkop1} for the scale factor on the brane. The function 
$C_{in}(t)$ can be determined by evaluating the solution at
$y=0$, and has the form
\be
C_{in}(t)=\pm \frac{\sqrt{2}A^2}{B^3(t)}\,
\log\Biggl(\frac{2}{B(t)}\,\Bigl[B^2(t) a_0(t)-A^2\Bigr]\Biggr)\,.
\ee
The remaining unknown function $E_{in}(t)$ will be determined from the 
continuity of the first derivative of $a(t,y)$ with respect to $y$
at the boundary $y=\Delta$ between the brane and the bulk universe. If
we use the form (\ref{solout}) for the solution outside the brane, this
condition can be written as
\be
\pm \sqrt{2} a_\Delta a_\Delta'=\sqrt{E_{in}(t) + B^2(t) a^2_\Delta
-2 A^2 a_\Delta}= \sqrt{E_{out}(t) + B^2(t) a^2_\Delta}
\ee
leading to the result
\be
E_{in}(t)= 2b^2\,\biggl[\,\frac{b^2}{\lambda^4}\,(\Delta-|y_{min}|)^2 -
\frac{a_\Delta^2}{\lambda^2} + \frac{2\hat{\kappa}^2}{3} 
\,\hat{\rho}_0 \,a_\Delta\,\biggr]\,.
\label{ein} 
\ee
Moreover, the function $E_{in}(t)$ is constrained by the (55)-component of
Einstein's equations. When the solution (\ref{sola}) is substituted in 
eq. (\ref{55}), we obtain the following constraint
\be
\frac{d E_{in}(t)}{dt}=\frac{4b^2}{3}\,\hat{\kappa}^2\,a^3 \dot{a}\,
(\hat{T}^5_5+\hat{\rho})\,.
\label{conein}
\ee
This is satisfied if, and only if, the fifth component of the energy-momentum
tensor, $\hat{T}^5_5$, on the brane is given by the expression in (\ref{55in}).

We have one more condition that our solution on the
brane needs to satisfy, namely, the vanishing of $a'(t,y)$ at the origin.
This leads to the result
\be
E_{in}(t)= 2A^2 a_0(t)-B^2(t) a^2_0(t)\,.
\label{dazero}
\ee
If we substitute the expression for $E_{in}(t)$ (\ref{ein}) in the above
equation, we find the following constraint
\be
\frac{b^2}{\lambda^2}\,(\Delta-|y_{min}|)^2 = \frac{(a_0-a_\Delta)}{\Delta}
\,\Bigl[2 a_0 |y_{min}|-\Delta\,(a_0+a_\Delta)\Bigr]\,,
\ee
where we have set $\hat{\rho}_0=\rho_0/2\Delta b$. 
In the limit $\Delta \rightarrow 0$, the above constraint should reduce
to the expression (\ref{minimum1}) for $|y_{min}|$. This condition allows us
to determine the function $a_\Delta(t)$ given in (\ref{adin}).

Finally, we extend the above analysis for the determination of the
thick-wall solution inside the brane, for a matter-dominated universe,
to the case of a brane-universe with a general equation of state
$p=w\,\rho$. In this case, the equation for the conservation of energy gives
the result $\hat{\rho}=\hat{\rho}_0/a^{3(1+w)}$. The first derivative
of the scale factor on the brane, now, takes the form
\be
a'(t,y)=\pm\,\frac{1}{\sqrt{2} a}\,\sqrt{E_{in}(t) + B^2(t) a^2(t,y)
-\frac{2}{(1-3w)}\,A^2 a(t,y)^{(1-3w)}}\,,
\label{difa}
\ee
and its continuity across the boundary $y=\Delta$ leads to the following
result
\be
E_{in}(t)= 2b^2\,\biggl[\,\frac{b^2}{\lambda^4}\,(\Delta-|y_{min}|)^2 -
\frac{a_\Delta^2}{\lambda^2} + \frac{2\hat{\kappa}^2}{3\,(1-3w)} 
\,\hat{\rho}_0 \,a_\Delta^{(1-3w)}\,\biggr]\,.
\label{einw} 
\ee
The constraint (\ref{conein}) that follows from the (55)-component of
Einstein's equations still holds and leads to the following expression for
the value of $\hat{T}^5_5$ on the brane
\be
\hat{T}^5_{5\,(br)}(t,y)=\hat{T}^5_{5\,(B)}(t,y) + \hat{\rho}\,
\Biggl[\frac{n(t,\Delta)}{n(t,y)}\,\Biggl(\frac{a(t,y)}{a(t,\Delta)}
\Biggr)^{3w}-1 \Biggr]
\ee
As in the previous case, we still have to fulfill the vanishing condition
of $a'(t,y)$ at the origin. Then, we find the result
\be
\frac{b^2}{\lambda^2}\,(\Delta-|y_{min}|)^2 = \frac{1}{\Delta}
\,\Biggl\{\frac{2 |y_{min}| a_0^2}{1-3w}\,\Bigl[1-
\Bigl(\frac{a_0}{a_\Delta}\Bigr)^{3w-1}\Biggr]-
\Delta\,(a_0^2-a_\Delta^2)\Biggr\}\,,
\ee
which, in the limit $\Delta \rightarrow 0$, reduces to
\be
\Bigl(\frac{a_\Delta}{a_0}\Bigr)^{1-3w}=
1-\frac{\hat{\kappa}^2}{12}\,\rho(t,0)\,\Delta b\,(1-3w)\,.
\ee
Note that for every value of $w$, larger or smaller that $1/3$, the
scale factor decreases as we move away from the origin, thus, ensuring
the monotonic behavior from the point $y=0$ to its minimum at
$y=|y_{min}|$. 
\par
However, the above expression gives a trivial result for the choice
$w=1/3$ which corresponds to a radiation-dominated universe. Moreover,
the expression (\ref{difa}) for the first derivative of $a(t,y)$ diverges for
the same choice. This means that we have to study this case separately. 
Starting all over again and setting $\hat{\rho}=\hat{\rho}_0/a^4$, the
first derivative, $a'(t,y)$, takes the form
\be
a'(t,y)=\pm\,\frac{1}{\sqrt{2} a}\,\sqrt{E_{in}(t) + B^2(t) a^2(t,y)
-2 A^2\,{\rm ln} a(t,y)}\,.
\ee
The result for $E_{in}(t)$ is accordingly modified and is given by
\be
E_{in}(t)= 2b^2\,\biggl[\,\frac{b^2}{\lambda^4}\,(\Delta-|y_{min}|)^2 -
\frac{a_\Delta^2}{\lambda^2} + \frac{2\hat{\kappa}^2}{3} 
\,\hat{\rho}_0 \,{\rm ln} a_\Delta\,\biggr]\,,
\label{einp} 
\ee
leading to the following form of $\hat{T}^5_5$ on the brane
\be
\hat{T}^5_{5\,(br)}=\hat{T}^5_{5\,(B)} + \hat{\rho}\,
\Biggl[\frac{n(t,\Delta)}{n(t,y)}\,\frac{a(t,y)}{a(t,\Delta)}-1 \Biggr]
\ee
Finally, the vanishing condition on $a'(t,y)$ at the origin leads to the 
constraint
\be
\frac{b^2}{\lambda^2}\,(\Delta-|y_{min}|)^2 = \frac{1}{\Delta}
\,\Biggl\{ 2 |y_{min}| a_0^2\,{\rm ln}\,\Bigl(\frac{a_0}{a_\Delta}\Bigr)
-\Delta\,(a_0^2-a_\Delta^2)\Biggr\}\,.
\ee
In the limit $\Delta \rightarrow 0$, we obtain the result
\be
a_\Delta(t)= a_0(t)\,{\rm exp}\Bigr(-\frac{\hat{\kappa}^2}{12}\,
\rho(t,0)\,\Delta b \Bigl)\,.
\ee
Once again, the scale factor on the brane is decreasing, as $y$ increases,
resulting in a similar behavior outside the brane until its minimum.


\end{document}